\documentclass[%
reprint,
nofootinbib,
amsmath,
amssymb,
prd,
longbibliography,
superscriptaddress]{revtex4-2}

\usepackage{mathrsfs}
\usepackage{graphicx}
\usepackage{dcolumn}
\usepackage{bm}
\usepackage{lipsum}
\usepackage{graphics}
\usepackage{bm}
\usepackage{bbm}
\usepackage{color}
\usepackage{microtype}
\usepackage{IEEEtrantools}
\usepackage{amsthm}
\usepackage{enumitem}
\usepackage[normalem]{ulem}
\usepackage{graphics}
\usepackage[T1]{fontenc}
\usepackage{pgfplots}
\usepackage{varwidth}

\usetikzlibrary{shapes, arrows, positioning}
\usepackage[T1]{fontenc}

\usepackage[capitalise]{cleveref}

\definecolor{darkgreen}{rgb}{0,0.5,0}

\def\GHPwt{\ensuremath{\circeq}}

\font\ec=ecrm0800 at 12pt
\def\thorn{\hbox{\ec\char'336}}
\def\edth{\hbox{\ec\char'360}}

\def\edthp{\hbox{\ec\char'360}'}

\def\mb{{\bar{m}}}

\def\l{\ell}
\def\m{\mathtt{m}}

\renewcommand{\S}{{\mathcal S}}

\newcommand{\T}{{\mathcal T}}

\renewcommand{\O}{\mathcal{O}}

\newcommand{\rhob}{{\bar{\rho}}}
\newcommand{\taub}{{\bar{\tau}}}

\def\GHPwt{\ensuremath{\circeq}}

\font\ec=ecrm0800 at 12pt
\def\thorn{\hbox{\ec\char'336}}
\def\edth{\hbox{\ec\char'360}}

\def\edthp{\hbox{\ec\char'360}'}

\def\mb{{\bar{m}}}

\def\mb{\bar{m}}

\def\GHPwt{\ensuremath{\circeq}}

\def\rhb{\bar\rho}

\def\tab{\bar\tau}

\def\th{\hbox{\ec\char'336}}
\def\edth{\hbox{\ec\char'360}}
\def\thp{\hbox{\ec\char'336}'}
\def\edthp{\hbox{\ec\char'360}'}

\def\thph{\tilde{\th}'}
\def\edthh{\tilde \edth}
\def\edthhp{\tilde \edthp}

\newcommand{\chand}[2]{\,{}_{#1} \mathcal{L}_{#2}{}}

\pgfplotsset{compat=1.18}
\begin{document}

\title{Analytically Separating the Source of the Teukolsky Equation }

\author{Andrew Spiers}
\affiliation{%
School of Mathematical Sciences and School of Physics and Astronomy, University of Nottingham, University Park, Nottingham NG7 2RD, United Kingdom; \\
 School of Mathematical Sciences and STAG Research Centre, University of Southampton, Southampton, SO17 1BJ, United Kingdom
}%

\date{\today}

\begin{abstract}
Recent gravitational wave detections from black hole mergers have underscored the critical role black hole perturbation theory and the Teukolsky equation play in understanding the behaviour of black holes. The separable nature of the Teukolsky equation has long been leveraged to study the vacuum linear Teukolsky equation; however, as theory and measurements advance, solving the sourced Teukolsky equation is becoming a frontier of research.  In particular, second-order calculations, such as in quasi-normal mode and self-force problems, have extended sources. This paper presents a novel method for analytically separating the Teukolsky equation's source, aimed to improve efficiency. Separating the source is a non-trivial problem due to the angular and radial mixing of generic quantities in Kerr spacetime. We provide a proof-of-concept demonstration of our method and show that it is accurate, separating the Teukolsky source produced by the stress-energy tensor of an ideal gas cloud surrounding a Kerr black hole. The detailed application of our method is provided in an accompanying \textit{Mathematica} notebook. Our approach opens up a new avenue for accurate black hole perturbation theory calculations with sources in both the time and frequency domain.

\end{abstract}

\maketitle


\section{\label{sec:Intro} Introduction}

Since the first detection of gravitational waves from a black hole binary~\cite{ligo} in 2015, gravitational wave astronomy has been a rapidly growing field. 
The current catalogue of binary detections~\cite{abbott2021population} is centred on binaries where the two compact objects have a similar mass. 
However, recently, the LIGO/Virgo/KAGRA collaboration detected a binary with mass-ratio of $\varepsilon=\frac{\mu}{M}\sim 0.04$~\cite{abbott2021gwtc}, where $M$ and $\mu$ are the masses of the primary the secondary binary component respectively. Gravitational astronomy is now not confined to the comparable mass regime; the so-called intermediate-mass-ratio inspirals (IMRIs, with mass ratio $10^{-2}  \gtrsim \varepsilon  \gtrsim 10^{-4}$) are increasingly more detectable~\cite{arca2020merging, sedda2021merging}. Additionally, millihertz frequency gravitational waves will be detectable with space-based gravitational wave interferometers, sensitive to lower mass ratios than ground-based detectors. In the 2030s, three space-based gravitational wave interferometer missions are planned: LISA~\cite{danzmann2017lisa}, TianQin~\cite{luo2016tianqin}, and Taiji~\cite{ruan2020taiji}.  Extreme-mass-ratio inspirals (EMRIs, $10^{-5}  \gtrsim \varepsilon  \gtrsim 10^{-8}$) are a key target for these missions~\cite{LISAConsortiumWaveformWorkingGroup:2023arg}. 


Numerical Relativity (NR)~\cite{baumgarte2010numerical} has had great success in accurately modelling comparable mass binaries. Progress is being made to extend the catalogue of NR waveforms into the IMRI regime~\cite{lousto2020exploring, rosato2021adapted}. Synergies with black hole perturbation theory can also improve efficiency~\cite{Wittek:2023nyi, Dhesi:2021yje}. However, these are challenging tasks due to the disparate length scales of such systems. In the EMRI regime, it is unrealistic to expect such methods to be capable of filling the parameter space of initial conditions in the LISA mission time frame. 

Alternatively, black hole perturbation theory provides an ideal method for approximating the binary spacetime in the small mass-ratio limit. The evolution of the binary can be modelled using the self-force approach~\cite{barackpound18}. In Refs.~\cite{pound2020second, wardell2021gravitational, warburton2021gravitational}, a two-timescale approximation~\cite{kevorkian1996method, hinderer2008two} was implemented for a self-force evolution of a binary, including (the dissipative) second-perturbative-order self-force. This model was limited to a binary of two Schwarzschild black holes in a quasi-circular inspiral. The waveforms were produced for a range of mass ratios and showed outstanding agreement with NR (before the inner-most stable circular orbit), even in the $\varepsilon\sim \frac{1}{10}$ regime. These results show that self-force waveforms have incredible potential for gravitational wave science. Crucially, the dissipative piece of the second-order self-force is a necessary contribution for high accuracy waveforms. There is also growing interest in second-order quasi-normal mode calculations~\cite{loutrel2021second, ripley2021numerical, ioka2007second, nakano2007second, cheung2023nonlinear, mitman2023nonlinearities, Lagos:2022otp,Redondo-Yuste:2023seq,Ma:2024qqnm}.

The recent second-order self-force results are limited to a Schwarzschild background black hole, whereas the primary black hole of astrophysical EMRIs is expected to have a significant spin ($a\sim 0.9M$). Adding the linear in-spin contributions perturbatively~\cite{MathewsPrimarySpin, mathews2022self} is straightforward. However, including the full nonlinear spin of the primary object is one of the outstanding problems in second-order self-force
. Working non-linearly in spin involves the background spacetime being a Kerr black hole~\cite{kerr1963gravitational}. The Kerr metric~\cite{kerr1963gravitational}, in Boyer--Lindquist coordinates~\cite{boyer1967maximal}, is
\begin{align}
    ds^2&= -\Big(1-\frac{2Mr}{\Sigma}\Big)dt^2+\frac{\Sigma}{\Delta}dr^2+\Sigma d\theta^2 \notag \\
    & \ \ \ +\Big( r^2 + a^2 +\frac{2Ma^2r\sin[\theta]}{\Sigma} \Big)\sin^2[\theta]d\phi^2 \notag \\
    & \ \ \ - \frac{4Mar\sin^2[\theta]}{\Sigma} dt d\phi, \label{eq:Kerr-metric}
\end{align}
where $M$ is the black hole mass, $a$ is the angular
momentum per unit mass, $\Sigma=r^2+a^2\cos^2[\theta]$, and $\Delta=r^2-2Mr+a^2$. Kerr spacetime is stationary and axially symmetric, but not spherically symmetric; hence, there is radial ($r$) and polar angle ($\theta$) mixing in the metric from~\eqref{eq:Kerr-metric}. This lack of symmetry makes the linearised Einstein field equations generally non-separable in Kerr; attempting to reduce the linearised Einstein field equations, a coupled set of PDEs, to a coupled set of ODEs by separation of variables seems fruitless. However, a separable field equation does exist, the Teukolsky equation~\cite{teuk1972, teuk1973}. Solving the sourced Teukolsky equation separably requires separating the angular and radial dependency in the source. In this paper, we produce a formalism for analytically separating general sources of the Teukolsky equation. Our method will be applied to produce accurate second-order self-force and quasi-normal mode calculations in follow-up papers. We expect our method will provide key efficiency savings in the second-order self-force problem, where a substantial amount of mode coupling is present. 

Before presenting our formalism, we briefly introduce black hole perturbation theory and the separability of the Teukolsky equation in the remainder of this section. In section~\ref{sec:decompmethod}, we present our method for decomposing the source (for a summary, see \cref{fig:Flow-chart}). In section~\ref{sec:IdealGasExample}, we apply our method to a toy example to show it accurately decomposes sources. Finally, in section~\ref{sec:Conclusion}, we make our concluding remarks. 

\subsection{Black hole perturbation theory and separability}

Black hole perturbation theory involves solving the Einstein field equations perturbatively around a background black hole spacetime. We write Einstein field equations in natural units as
\begin{align}\label{eq:EFE}
G_{ab}[g_{cd}]=8\pi T_{ab},
\end{align}
where $G_{ab}$ is the Einstein tensor, $g_{ab}$ is the metric\footnote{The indices on the metric in \cref{eq:EFE} indicate that the metric is a rank two tensor, rather than being free indices in the equation (such as those on the Einstein operator and stress-energy tensor). Throughout this work, indices that appear on tensors inside square brackets are there for labelling purposes only.}, and $T_{ab}$ is the stress-energy tensor. The metric and stress-energy tensor are expanded in orders of the small parameter $\varepsilon$,
\begin{align}\label{eq:metricexpansion}
g_{ab} &= g^{(0)}_{ab}+\varepsilon h^{(1)}_{ab} +\varepsilon^2 h^{(2)}_{ab} +\ldots+ \varepsilon^n h^{(n)}_{ab}+\ldots,\\
T_{ab} &= T^{(0)}_{ab}+\varepsilon T^{(1)}_{ab} +\varepsilon^2 T^{(2)}_{ab} +\ldots+ \varepsilon^n T^{(n)}_{ab}+\ldots\label{eq:stressexpansion}
\end{align}
We take $g^{(0)}_{ab}$, the background metric, to be the Kerr metric. Therefore, $T^{(0)}_{ab}=0$, as Kerr is a vacuum solution. Under the expansions in \cref{eq:metricexpansion,eq:stressexpansion}, the Einstein field equation, \cref{eq:EFE}, can be expressed as a hierarchical set of linear field equations in ascending order of $\varepsilon$~\cite{wardell2021gravitational, mySecond-orderTeuk},
\begin{align} 
\delta G_{ab}[h^{(1)}_{ab}]&=8\pi T^{(1)}_{ab},\label{eq:linearEFE}\\
\delta G_{ab}[h^{(2)}_{ab}]&=8\pi T^{(2)}_{ab}-\delta^2G_{ab}[h^{(1)}_{ab},h^{(1)}_{ab}],\label{eq:quadraticEFE}\\
\delta G_{ab}[h^{(3)}_{ab}]&=\ldots
\end{align}
where $\delta G_{ab}$ is the linearised Einstein operator and $\delta^2 G_{ab}$ is the quadratic Einstein operator~\cite{mySecond-orderTeuk}. 

The abundant symmetry of Schwarzschild spacetime makes the linearized Einstein operator, $\delta G_{ab}$, separable~\cite{martel2005gravitational, Spiers:2023Coupling}. However, the linearized Einstein operator in Kerr spacetime is generally non-separable. It is, therefore, remarkable that separable field equations exist, the Teukolsky equations~\cite{teuk1972, teuk1973},
\begin{align}\label{eq:teuk}
    \O \psi_0^{(1)} = \S_0 T^{(1)}_{ab}, \ \ \ \ \O^\prime \psi^{(1)}_4 = \S_4 T^{(1)}_{ab},
\end{align}
where $\psi_0^{(1)}=\T_0 h^{(1)}_{ab}$, $\psi_4^{(1)}=\T_4 h^{(1)}_{ab}$, and $\T_0$, $\T_4$, $\S_0$, $\S_4$, $\O$, and $\O^\prime$ are second order differential operators, see Ref.~\cite{wardellpoundreview2021}.

It is not immediately obvious that \cref{eq:teuk} is separable and relates to Kerr spacetime being Petrov type-D~\cite{petrov2000classification} and vacuum\footnote{Ref~\cite{dudley1977separation} showed the Teukolsky equations of general vacuum Petrov type-D spacetimes are separable.}. To show \cref{eq:teuk} is separable, one can choose a tetrad and convert to master Teukolsky equation form~\cite{teuk1972, teuk1973, wardellpoundreview2021}. In the Kinnersley tetrad~\cite{kinnersley1969type}, the master Teukolsky equations are
\begin{align}\label{eq:masterTeuk}
\hat{\mathcal{O}}_{+2}[\psi_0^{(1)}]&=-\Sigma \mathcal{S}_0[8\pi T^{(1)}_{ab}], \\
\hat{\mathcal{O}}_{-2}[\zeta^{4}\psi_4^{(1)}]&=-\Sigma \zeta^{4}\mathcal{S}_4[8\pi T^{(1)}_{ab}],
\end{align}
where $\hat{\mathcal{O}}_{s}$ is the spin $s$ master Teukolsky operator and
\begin{align}
    \zeta:= r-ia\cos[\theta].\label{eq:zeta}
\end{align}
The separability ansatz relies on the time dependence of the Teukolsky master variable, $\psi_0^{(1)}$ or $\zeta^{4}\psi_4^{(1)}$, being expressed as proportional to $e^{-i\omega t}$ (where $\omega$ may be complex); for example, working in the frequency domain. The spin $s$ master Teukolsky operator can then be expressed as
\begin{align}
    \hat{\mathcal{O}}_{s}=\mathcal{R}_{s}+ \Theta_{s},
\end{align}
where~\cite{wardellpoundreview2021}, in Boyer--Lindquist coordinates,
\begin{align}
    \mathcal{R}_{s}&= \bigg( \Delta^{-s} \frac{d}{dr}\Big(\Delta^{s+1} \frac{d}{dr}\Big) + \frac{K^2-2is(r-M)K}{\Delta} \notag \\
    & \ \ \ + 4is\omega r -{}_s \lambda_{\l\m} \bigg),
\end{align}
where $K:=(r^2+a^2)\omega-a\m$, and 
\begin{align}
    \Theta_{s}&=\bigg( \frac{d}{d\chi}\Big((1-\chi^2)\frac{d}{d\chi}\Big) + a^2\omega^2\chi^2 - \frac{(\m+s\chi)}{1-\chi^2} \notag \\
     & \ \ \ -2as\omega\chi +s+A \bigg),
\end{align}
where $\chi:=\cos[\theta]$ and $A:= {}_s \lambda_{\l\m} +2a\m\omega -a^2\omega^2$ (and the eigenvalues ${}_s \lambda_{\l\m}$ depend on $a\omega$). The eigenfunctions of $\Theta_{s}$ are spin-weighted spheroidal harmonics\footnote{In the limit $a\omega\rightarrow 0$ the spin-weighted spheroidal harmonics reduce to spin-weighted spherical harmonics.}~\cite{teuk1972, teuk1973}, ${}_s S_{\l\m}[\theta,\phi,a\omega]$; that is,
\begin{align}
    \Theta_{s} \big[ {}_s S_{\l\m}\big] =0 .
\end{align}

More generally, the Teukolsky equation can be shown to be separable in any principle-null direction aligned tetrad in Kerr without using coordinates~\cite{aksteiner2014geometry, wardellpoundreview2021}. The Teukolsky equation can be written as
\begin{align}
    \zeta\bar{\zeta}\O=\mathfrak{R} -\mathfrak{S} ,
\end{align}
where $\mathfrak{R}$ and $\mathfrak{S}$ are operators that commute; hence, the equation is separable. 

Research has largely focused on solving the first-order vacuum Teukolsky equation~\cite{Kokkotas:1999bd,Berti:2009kk,Konoplya:2011qq} or point-particle source problem~\cite{barackpound18, wardellpoundreview2021}. Hence, previous research has generally not required efficient separation of the right-hand side of the Teukolsky equation into modes of the separable Teukolsky equation.  As calculations move to second order, sources are unavoidable. The second-order Teukolsky equation comes in two forms, the Campanelli--Lousto second-order Teukolsky equation~\cite{l-c} and the reduced-second-order Teukolsky equation~\cite{green2020teukolsky, mySecond-orderTeuk}. Both equations are sourced by the stress-energy perturbations and a quadratic operator acting on the first-order metric perturbation. For example, the reduced second-order Teukolsky equation takes the form
\begin{align}
    \mathcal{O}[\psi_{0L}^{(2)}]&=\mathcal{S}_0\big[8\pi T^{(2)}_{ab} - \delta^2 G_{ab}[h^{(1)}_{ab},h^{(1)}_{ab}]\big], \label{eq:reducedTeuk0}\\
    \mathcal{O}^\prime[\psi_{4L}^{(2)}]&=\mathcal{S}_4\big[8\pi T^{(2)}_{ab} - \delta^2 G_{ab}[h^{(1)}_{ab},h^{(1)}_{ab}]\big],\label{eq:reducedTeuk4}
\end{align}
where $\psi_{0L}^{(2)}=\T_0 [h^{(2)}_{ab}]$, $\psi_{4L}^{(2)}=\T_4 [h^{(2)}_{ab}]$. Note that $\psi_{0L}^{(2)}$ and $\psi_{4L}^{(2)}$ are the linear in $h^{(2)}_{ab}$ part of the full second-order Weyl scalars,
\begin{align}
    \psi_{0}^{(2)}&= \psi_{0L}^{(2)} +\psi_{0Q}^{(2)},\\
    \psi_{4}^{(2)}&= \psi_{4L}^{(2)} +\psi_{4Q}^{(2)},
\end{align}
where $\psi_{0Q}^{(2)}=\delta^2\psi_0[h^{(1)}_{ab},h^{(1)}_{ab}]$ and $\psi_{4Q}^{(2)}=\delta^2\psi_4[h^{(1)}_{ab},h^{(1)}_{ab}]$ are the quadratic in $h^{(1)}_{ab}$ parts of $\psi_{0}^{(2)}$ and $\psi_{4}^{(2)}$ respectively~\cite{l-c, mySecond-orderTeuk}; the operators $\delta^2\psi_0$ and $\delta^2\psi_4$ are given in Ref.~\cite{2nd-order-notebook}. The $\delta^2 G_{ab}[h^{(1)}_{ab},h^{(1)}_{ab}]$ parts of \cref{eq:reducedTeuk0,eq:reducedTeuk4} make the source non-compact, extending from the horizon of the primary to future null infinity. 

We need a precise definition for separating the source of the Teukolsky equation. Take the master Teukolsky equation with a generic source $f[t,r,\theta,\phi]$,
\begin{align}
    \hat{\mathcal{O}}_{s}[\psi]=f,\label{eq:sourcedTeukolsky}
\end{align}
for the spin-$s$ master Teukolsky variable $\psi$. In order to solve the equation separably, $\psi$ and $f$ must be expressed in a mode decomposition form:
\begin{align}\label{eq:phidecomp}
    \psi[t,r,\theta,\phi]= \sum_{\l=|s|}^\infty \sum_{\m=-l}^{l} \psi_{\l\m}[r]e^{-i\omega t} {}_s S_{\l\m}[\theta,\phi],  \\
    f[t,r,\theta,\phi]= \sum_{\l=|s|}^\infty \sum_{\m=-l}^{l} f_{\l\m}[r]e^{-i\omega t} {}_s S_{\l\m}[\theta,\phi].\label{eq:fdecomp}
\end{align}
Then, \cref{eq:sourcedTeukolsky} reduces to 
\begin{align}\label{eq:seperableTeuk}
    \mathcal{R}_{s}[\psi_{\l\m}]=f_{\l\m}
\end{align}
for each $\l$ and $\m$.

As spin-weighted spheroidal-harmonics are an orthonormal set~\cite{breuer1977some} of eigenvectors on $S^2$, one can separate the source of the Teukolsky equation of spin weight $s$, using
\begin{align}\label{eq:Soint}
    f_{\l\m}[r]e^{-i\omega t} =\oint {}_s S^*_{\l\m} f[t,r,\theta,\phi] d\Omega,
\end{align}
where ${}_s S^*_{\l\m}$ is the bi-orthogonal dual~\cite{london2023biorthogonal} of the spin-weighted spheroidal harmonic which simplifies to the complex conjugate when $\omega$ is real. $f_{\l\m}[t,r]$ in \cref{eq:Soint} will then satisfy \cref{eq:fdecomp}. As spin-weighted spheroidal harmonics generally have no known closed form, integrating Eq.~\eqref{eq:Soint} is challenging. One could integrate numerically, but $f[t,r,\theta,\phi]$ is not generally separable in the radial and polar angle coordinates. Therefore, the integral will need to be computed numerically at each radial point on a grid for each $\l$ and $\m$ mode. This method has been used to separate the source of the Teukolsky equation in Ref.~\cite{Ma:2024qqnm} for a quadratic quasi-normal mode calculation. However, the inefficiency of the numerical integrals may be problematic for self-force calculations where multiple modes must be calculated, and second-order calculations must fill a four-dimensional parameter space of initial data to create the waveform templates for LISA data analysis.

Alternatively, one could avoid separating the source of the Teukolsky equation by solving the Teukolsky equation in PDE form~\cite{harms2013numerical, PanossoMacedo:2019npm}. However, numerical methods for solving PDEs convert the PDE into a system of ODEs. Hence, it is likely that leveraging the separability of the Teukolsky equation will develop the most efficient algorithm for separating the Teukolsky equation into ODEs.

\section{Decomposing General functions in Kerr spacetime}\label{sec:decompmethod}

Our formalism is designed to separate general sources in Kerr into the sum of the product of radial functions and spin-weighted spheroidal harmonics. Essentially, we separate the radial and polar angle dependency\footnote{Separating the time and azimuthal angle dependencies is trivial due to the symmetries of Kerr spacetime.} and express the angular dependence in terms of spin-weighted spheroidal harmonics.

An immediate simplification we can make is re-expanding the spheroidal harmonics into spin-weighted spherical harmonics. Spin-weighted spherical harmonics are an easier basis of functions to work with because they are closed-form and have simple spin-weight raising and lowering operators associated with them~\cite{goldberg1967spin}\footnote{Spin-weight raising and lowering operators for the spin-weighted spheroidal harmonics have been derived~\cite{shah2016raising} but as an expansion in orders of $a\omega$.}. Leavers method~\cite{leaver1986solutions} is the most accurate method of expressing spin-weighted spheroidal harmonics but does not expand them in terms of spin-weighted spherical harmonics directly. Press and Teukolsky~\cite{press1973perturbations} were the first to expand spin-weighted spheroidal harmonics into spin-weighted spherical harmonics. Ref.~\cite{hughes2000evolution} then found a more efficient method, which has been implemented in the Black Hole Perturbation Toolkit~\cite{BHPToolkit}. The expansion in terms of spherical harmonics converges to Leaver's method solutions and is invertible. But, to work with either expansion, one must truncate at a finite order. Hence, for the remainder of this paper, we assume any inputs that are generally expressed in terms of spin-weighted spheroidal harmonics can be re-expanded in terms of spin-weighted spherical harmonics. Similarly, it is sufficient to express the source in terms of spin-weighted spherical harmonics as we can re-expand it in terms of spin-weighted spheroidal harmonics. That is, we can simplify our goal to expressing the source as
\begin{align}\label{eq:sourceY}
    f[t,r,\theta,\phi]= \sum_{\l=|s|}^\infty \sum_{\m=-\l}^{\l} \hat f_{\l\m}[r]e^{-i\omega t} {}_s Y_{\l\m}[\theta,\phi].
\end{align}
where ${}_s Y_{\l\m}[\theta,\phi]$ are the spin-weighted spherical harmonics.

The rest of this section is dedicated to expressing the source of the Teukolsky equation in the form of Eq.~\eqref{eq:sourceY}. To simplify this task, we use the Newman--Penrose~\cite{np1962}, Geroch--Held--Penrose (GHP)~\cite{geroch1973space}, and Held~\cite{held1974formalism} formalisms. We also find there is a preferred tetrad and coordinate scheme to express the source in Eq.~\eqref{eq:sourceY} form. In such a tetrad and coordinates, almost all the Kerr background quantities are naturally expressed as single spin-weighted spherical harmonics, and all angular derivatives become spin-weight raising/lowering operators. One problematic background quantity is unavoidable, $\rho$ (and $\bar\rho$), where
\begin{align}\label{eq:rho}
    \rho=-\frac{1}{\zeta}=\frac{-1}{(r-ia\cos[\theta])}.
\end{align}
As the radial and angular dependence appears on the denominator, $\rho$ does not naturally separate into a radial and angular function. In Sec.~\ref{sec:rhoexpansions}, we separate the radial and angular dependence in Eq.~\eqref{eq:rho} using a Fourier expansion. Combining these methods, we produce a complete formalism for expressing the source as a convergent sum of spin-weighted spherical harmonics (Eq.~\eqref{eq:sourceY}).

\subsection{Spin-weighted spherical harmonics}

The spin-weighted spherical harmonics were defined in Ref.~\cite{goldberg1967spin}, and we follow their conventions up to a minus sign in the definition of the spin raising and lowering operators (similarly to Ref.~\cite{Spiers:2023Coupling}),
\begin{align}\label{eq:altedth}
\hat{\edth}&= (\sin[\theta])^{-s}(\partial_\theta +i \csc[\theta] \partial_\phi)(\sin[\theta])^{s} \notag \\
&=\partial_\theta +i\csc[\theta]\partial_\phi -s\cot[\theta], \\
\hat{\edthp}&= (\sin[\theta])^{-s}(\partial_\theta -i \csc[\theta] \partial_\phi)(\sin[\theta])^{s} \notag \\
&=\partial_\theta -i\csc[\theta]\partial_\phi -s\cot[\theta],
\end{align}
when acting on a spin-$s$ weighted quantity. One can define the spin-$s$ spherical harmonics from the scalar harmonics,
\begin{equation}
  {}_s Y_{\l\m}:=
    \begin{cases}
      &\sqrt{\frac{(l-|s|)!}{(l+|s|)!}}(-1)^s\hat\edth^s Y_{\l\m}, \ \ 0\leq s \leq \l,\\
      & \sqrt{\frac{(l-|s|)!}{(l+|s|)!}}\hat\edthp^{|s|} Y_{\l\m}, \   \ \ \ \  -\l\leq s \leq 0.
    \end{cases}       
\end{equation}
The following relations then hold~\cite{goldberg1967spin},
\begin{align}
\hat{\edth} \  {}_{s}Y_{\l,\m} =&-\sqrt{(l-s)(l+s+1)} \  {}_{s+1}Y_{\l,\m}, \label{eq:ethswshdef} \\
\hat{\edth}^\prime \  {}_{s}Y_{\l,\m} =&\sqrt{(l+s)(l-s+1)} \  {}_{s-1}Y_{\l,\m}, \label{eq:ethprimeswshdef} \\
\hat{\edth}^\prime \hat{\edth} \  {}_{s}Y_{\l,\m} =&-(l-s)(l+s+1) \  {}_{s}Y_{\l,\m} ,\label{eq:Laplacedef} \\
{}_s\overline{Y}_{\l,\m}=&(-1)^{s+\m}\ {}_{-s}Y_{l,-\m}. \label{eq:swsvhbardef}
\end{align}

The product of two spin-weighted spherical harmonics can be expressed as the sum of spin-weighted spherical harmonics,
\begin{align}\label{eq:reexpandY}
    {}_{s_1}Y_{\l_1,\m_1}{}_{s_2}Y_{\l_2,\m_2} =\sum_{\l_1=s_1,\m_1}^\infty \sum_{\l_2=s_2,\m_2}^\infty C^{\l,\m,s}_{\l_1,\m_1,s_1,\l_2,\m_2,s_2}{}_{s}Y_{\l,\m}.
\end{align}
where $C^{\l,\m,s}_{\l_1,\m_1,s_1,\l_2,\m_2,s_2}$ is equivalent to a surface integral,
\begin{align}\label{eq:Cintegral}
    C^{\l,\m,s}_{\l_1,\m_1,s_1,\l_2,\m_2,s_2}=\oint {}_{s}\bar{Y}_{\l,\m}\ {}_{s_1}Y_{\l_1,\m_1}\ {}_{s_2}Y_{\l_2,\m_2}.
\end{align}
\noindent As spin-weighted spherical harmonics are related to Wigner-D matrices, \cref{eq:Cintegral} can be evaluated algebraically using 3j symbols~\cite{hecht2012quantum, shiraishi2013probing, Spiers:2023Coupling},
\begin{align}\label{coupling}
C^{ \l \m s}_{ \l'\m's' \l''\m''s''} &= (-1)^{\m+s}\sqrt{\frac{(2 \l+1)(2 \l'+1)(2 \l''+1)}{4\pi}}\times \notag \\
& \ \ \ \begin{pmatrix} \l &  \l' &  \l'' \\ s & -s' & -s''\end{pmatrix}
\begin{pmatrix} \l &  \l' &  \l'' \\ -\m & \m' & \m''\end{pmatrix}\!.
\end{align}

\subsection{The Newman--Penrose formalism}\label{sec:NP}

The Newman--Penrose (NP) formalism~\cite{np1962} utilises a basis of four null vectors, called a \textit{tetrad}, to express curvature quantities. 
The NP basis vectors are labelled 
\begin{align}
    e_{[a]}^a:=\{e_{[1]}^a,e_{[2]}^a,e_{[3]}^a,e_{[4]}^a\} := \{l^a,n^a,m^a,\bar{m}^a\},
\end{align}
where indices in square brackets are tetrad indices. The vectors $l^a$ and $n^a$ are real, and $m^a$ is complex. Overbars denote a complex conjugation. Conventionally, for a positive metric signature, the orthonormal relationship of the tetrad takes the form
\begin{equation}\label{eq:orthonormal}
l^an_a=-1,\quad  m^a\bar{m}_a=1.
\end{equation}
As the tetrad is orthonormal, following Eq.~\eqref{eq:orthonormal}, the metric can be expressed as~\cite{np1962},
\begin{equation}
  g_{ab} = -2 l_{(a} n_{b)} +2 m_{(a} \bar{m}_{b)} .
\end{equation}

The NP formalism uses Ricci rotation coefficients to express the connection~\cite{waldbook},
\begin{equation}
\gamma_{[c][a][b]}=e_{[c]}^ke_{[a]k;i}e_{[b]}^i.\label{eq:RRC}
\end{equation}
There are 24 independent components of Ricci rotation coefficients. In the NP formalism, the components are expressed as 12 complex scalars, known as spin coefficients:
\begin{align}
\kappa=-\gamma_{[3][1][1]},\ \tau=-&\gamma_{[3][1][2]},  \ \sigma=-\gamma_{[3][1][3]}, \notag
\\ \notag
\\  \
\rho=-\gamma_{[3][1][4]}, \pi=-&\gamma_{[2][4][1]},\ \nu=-\gamma_{[2][4][2]}, \notag
\\ \notag
\\  \ \mu=-\gamma_{[2][4][3]},& \ \lambda=-\gamma_{[2][4][4]}, \notag
\\ \notag
\\ 
\epsilon=-\frac{\gamma_{[2][1][1]}+\gamma_{[3][4][1]}}{2}, & \ \gamma=-\frac{\gamma_{[2][1][2]}+\gamma_{[3][4][2]}}{2}, \notag
\\ \notag
\\ 
\beta=-\frac{\gamma_{[2][1][3]}+\gamma_{[3][4][3]}}{2}, & \ \alpha=-\frac{\gamma_{[2][1][4]}+\gamma_{[3][4][4]}}{2}. \label{eq:spincoeffs}
\end{align}

In vacuum spacetimes, the Ricci curvature is zero. The Weyl tensor contains the vacuum curvature. In the NP formalism, the ten degrees of freedom of the Weyl tensor are expressed using five complex scalars, known as Weyl scalars:
\begin{align}
\psi_0 =& C_{[1][3][1][3]}, \\
\psi_1 =& C_{[1][3][1][2]},  \\
\psi_2 =& C_{[1][3][4][2]},  \\
\psi_3 =& C_{[1][2][4][2]},  \\
\psi_4 =& C_{[2][4][2][4]}  .
\end{align}

Kerr spacetime has two principal null vectors (two pairs of principal null vectors which coincide)~\cite{petrov2000classification, chandrabook}. That is, Kerr is Petrov type D. A tetrad can be chosen such that $l^a$ and $n^a$ are tangent to the principal null directions and four of the Weyl scalars and four spin coefficients then vanish \cite{goldberg2009republication, chandrabook},
\begin{align}\label{petrovconds1}
&\psi_0=0,\psi_1=0,\psi_3=0,\psi_4=0, \\
&\kappa=0, \lambda=0, \nu=0, \sigma=0.\label{petrovconds2}
\end{align}
These simplifications were used to help derive the Teukolsky equation~\cite{teuk1973}.

The covariant derivative is also expressed using tetrad components in the NP formalism:
\begin{align}
&D\eta:=\eta_{|[1]}:=l^a \nabla_a\eta , \ \ \ \ \Delta\eta:=\eta_{|[2]}:=n^a \nabla_a\eta, \notag \\  &\delta\eta:=\eta_{|[3]}:=m^a \nabla_a\eta, \ \ \ \  \overline{\delta}\eta:=\eta_{|[4]}:=\overline{m}^a \nabla_a\eta.               \label{paradervs}
\end{align}

\subsection{The GHP formalism}\label{subsec:GHP}

The GHP formalism builds on the NP formalism, helping to represent the symmetry in principal null direction aligned tetrads in Petrov type D spacetimes (such as the Kinnersley~\cite{kinnersley1969type}, Carter~\cite{carter1968global}, and Hartle--Hawking~\cite{hawking1972energy} tetrads). While constraining $l^a$ and $n^a$ to point in the same direction, two tetrad rotation degrees of freedom remain unconstrained. These freedoms can be associated with spin and boost transformations and are isomorphic to the group of multiplication by a complex number, $\vartheta$~\cite{toomani2021new}.  To express the weight of a GHP quantity $f$ we use the notation $f \GHPwt \{p,q\}$, meaning that under a spin and boost transformation
  \begin{equation}
      f \xrightarrow{} \vartheta^p \bar{\vartheta}^q f \ \ \Longleftrightarrow \ \  f \GHPwt \{p,q\}.
  \end{equation}

\noindent $p$ and $q$ are known as GHP weights that can be equated to the spin ($s=\frac{1}{2}(p-q)$) and boost ($b=\frac{1}{2}(p+q)$) weights. Products of two tensors with weights ${a,b}$ and ${c,d}$ produces a tensor of weight ${a+c,b+d}$. On the other hand, the addition of two quantities can only be performed if they are of the same weight.

The $[b,s]$ weights of the tetrad vectors are, $[1,0]$, $[-1,0]$, $[0,1]$, and $[0,-1]$ for $l^a$, $n^a$, $m^a$, and $\bar{m}^a$ respectively. Similarly, the $\{p,q\}$ weights of the tetrad vectors are $\{1,1\}$, $\{-1,-1\}$, $\{1,-1\}$, and $\{-1,1\}$ respectively. 

In Petrov type-D principle null direction aligned tetrads, there is a freedom to interchange $l^a$ and $n^a$. GHP introduced a prime operation to represent the interchange $l^a\rightarrow n^a$, $n^a\rightarrow l^a$, $m^a\rightarrow \mb^a$, and $\mb^a\rightarrow m^a$. The prime operation affects the GHP weights accordingly, $f' \circeq \{-p,-q\}$. Complex conjugation also affects the GHP weights: $\bar{f}\circeq \{q,p\}$.

In the GHP formalism, half of the NP spin coefficients are relabelled using the prime operation notation\footnote{For example, using \cref{eq:spincoeffs}, $\kappa^\prime=-\gamma_{[4][2][2]}=-\nu$ where we have used the identity $\gamma_{[c][a][b]}=-\gamma_{[c][b][a]}$.},
\begin{align}
\kappa^\prime:=-\nu, \quad \sigma^\prime:=-\lambda, \quad \rho^\prime:=-\mu, \notag \\ \tau^\prime:=-\pi, \quad \beta^\prime:=-\alpha, \quad \epsilon^\prime:=-\gamma.
\end{align}
\noindent The GHP weights of the spin coefficients (and their primes) follow directly from the weights of the tetrad vectors (using Eqs.~\eqref{eq:spincoeffs})\footnote{For example, $\kappa= -m^k l_{k,i} l^i\circeq\{3,1\}$, as $m^a\circeq\{1,-1\}$ and $l^a\circeq\{1,1\}$.}; that is,
\begin{align}
\kappa\circeq\{3,1\},  \ \sigma\circeq\{3,-1\}, \ \rho\circeq\{1,1\},  \tau\circeq\{1,-1\}.
\end{align}
\noindent The spin coefficients $\epsilon$, $\epsilon^\prime$, $\beta$, and $\beta^\prime$ do not have well-defined weights. Similarly, the NP derivative operators do not have well-defined weights. GHP found by combining these poorly defined weight quantities, one can produce derivative operators with well-defined weights,
\begin{align}
\thorn \eta = (D-p\epsilon-q\bar{\epsilon})\eta , \ \thorn^\prime \eta = (\Delta+p\epsilon^\prime+q\bar{\epsilon}^\prime)\eta, \notag \\
\edth \eta = (\delta-p\beta+q\bar{\beta}^\prime)\eta , \ \edth^\prime \eta = (\bar{\delta}+p\beta^\prime-q\bar{\beta})\eta, \label{eq:GHPderivatives} 
\end{align}
where $\eta\GHPwt\{p,q\}$. Equation~\eqref{eq:GHPderivatives} respectively have boost and spin weights of $[1,0]$, $[-1,0]$, $[0,1]$, and $[0,-1]$ and GHP weights of $\{1,1\}$, $\{-1,-1\}$, $\{1,-1\}$, and $\{-1,1\}$, plus the weights of $\eta$.

Two clear advantages of the GHP formalism are that the equations are more condensed than in NP form, and they offer a straightforward consistency check by checking that the weights of an equation are consistent.

We will also make use of the GHP commutation relations~\cite{geroch1973space} (with \cref{petrovconds1,petrovconds2} imposed),
\begin{align}
    \thp\th\eta&=\th\thp\eta-\eta\big(p(\psi_2+\tau\tau')+q(\taub'\taub-\bar\psi_2)\big) \notag \\
    & \ \ \ -(\taub-\tau')\edth\eta -(\tau-\bar\tau')\edthp\eta,\label{eq:GHPCommutation1} \\
    \edth\th\eta&=\th\edth\eta-q\eta \rhob\taub'+\taub'\th\eta-\rhob\edth\eta, \label{eq:GHPCommutation2}\\
    \edth\thp\eta&=\thp\edth\eta+p\rho'\tau+\tau\thp\eta-\rho'\edth\eta.\label{eq:GHPCommutation3}
\end{align}
We apply the commutation relations, \cref{eq:GHPCommutation1,eq:GHPCommutation2,eq:GHPCommutation3} (and their complex conjugates) to commute all $\edth$ and $\edthp$ derivatives to the right, and all $\thp$ derivatives to the right of $\th$. This simplifies equations, and this ordering will place all spin-raising and lowering operators to the right of radial derivatives in section~\ref{sec:coords}.

\subsection{The Held Formalism}\label{sec:Held-Formalism}

The Held formalism builds on the GHP formalism, producing an algorithmic integration technique for a section of the GHP equations (the Ricci and Bianchi identities in GHP form). The Held integration method has recently been applied to solve black hole perturbation theory problems~\cite{green2020teukolsky,toomani2021new,price2007existence}. Here, we review the main aspects of the Held formalism and its application in the Kinnersly tetrad~\cite{kinnersley1969type}. In Boyer--Lindquist coordinates, the Kinnersley tetrad is
\begin{align}
    l^a&=\frac{1}{\Delta}\Big\{   r^2+a^2, \Delta, 0 , a  \Big\},\label{eq:Kinnersley-l}  \\
    n^a&=\frac{1}{2\Sigma}\Big\{   r^2+a^2, -\Delta, 0 , a  \Big\},\label{eq:Kinnersley-n} \\
    m^a&=\frac{1}{\sqrt{2}\bar\zeta}\Big\{   ia\sin[\theta], 0, 1 , \frac{i}{\sin[\theta]}  \Big\}, \label{eq:Kinnersley-m}\\
    \bar m^a&=\frac{1}{\sqrt{2}\zeta}\Big\{   -ia\sin[\theta], 0, 1 , \frac{-i}{\sin[\theta]}  \Big\},    \label{eq:Kinnersley-mbar} 
\end{align}
I also comment on the advantages of the Held formalism and when the integration technique can be used.

The Held formalism leverages the Ricci identities (called ``Field equations'' in~\cite{held1974formalism}), Bianchi identities, and commutation relations to build a set of equations for derivatives of background Kerr quantities. For example, in a vacuum Petrov type-D spacetime, the $R_{[1][3][1][4]}$ Ricci identity~\cite{chandrabook} gives
\begin{align}\label{eq:thornrho}
\th \rho &= \rho^2.
\end{align}
Equation~\eqref{eq:thornrho} can be interpreted as the integral of $\th\rho$. Using this identity to solve differential equations involving $\rho$ and $\th$ is known as Held integration~\cite{held1974formalism}. For example, one may solve the complex ordinary differential equation.
\begin{align}\label{eq:Heldexmple}
    \th A[\rho] = B^\circ \rho^3,
\end{align}
\noindent where $A[\rho]$ is an unknown function of $\rho$ and $B^\circ$ is a function independent of $\rho$ ($\th B^\circ=0$, all variables labelled with $\circ$ superscripts share this property and will be referred to as Held scalars\footnote{While quantities labelled with a superscript $\circ$ are commonly known as Held scalars, the notation was originally introduced by Kinnersley~\cite{kinnersley1969type}.}). Using \cref{eq:thornrho} one can find the solution, 
\begin{align}\label{eq:HeldexmpleAnswer}
    A[\rho] = \frac{B^\circ \rho^2}{2} + C^\circ,
\end{align}
\noindent where $C^\circ$ is any function independent of $\rho$.

However, Held integration does not apply to general GHP equations. When an alternative derivative operator (such as $\thp$, $\edth$, or $\edthp$) acts on an unknown variable in a differential equation, the Held method of integration is not helpful. For example, 
\begin{align}
    \edth D[\rho] = E^\circ \rho^3,\label{eq:edthD}
\end{align}
cannot be solved using Held integration. Within these limitations, the Held integration method is useful for solving certain problems in black hole perturbation theory~\cite{green2020teukolsky,toomani2021new,price2007existence}. 

To apply Held integration to more generic equations than Eq.~\eqref{eq:Heldexmple}, such as when GHP derivatives act on spin coefficients, the Held formalism extracts the $\rho$ dependency out of such quantities. Held began by modifying the derivative operators of GHP. The Held derivative operators are $\th$, $\thph$, $\edthh$, and $\edthhp$. The new operators are defined as~\cite{green2020teukolsky, toomani2021new},
\begin{align}
    \thph &=\th' - \tab \edth - \tau \edthp + \tau \tab \left( \frac{p}{\bar \rho} + \frac{q}{\rho} \right) + \frac{1}{2} \left( \frac{q \bar{\Psi}_2}{\bar{\rho}} + \frac{p \Psi_2}{\rho} \right)  \label{eq:thph} , \\
    \edthh &=\frac{1}{\bar \rho} \edth + \frac{q\tau}{\rho}  ,\label{eq:edthh} \\
    \edthhp &= \frac{1}{\rho} \edth' + \frac{p\tab}{\bar \rho} \label{eq:edthph},
\end{align}
\noindent with GHP weights $\{ -1,-1 \}$, $\{ 0,-2 \}$, and $\{ -2,0 \}$ respectively. These derivative operators were chosen to satisfy the following commutation relations~\cite{held1974formalism},
\begin{align}\label{eq:HeldCom}
    \Big[\th, \thph\Big]\eta^\circ =\Big[\th, \edthh\Big]\eta^\circ=\Big[\th, \edthhp\Big]\eta^\circ=0,
\end{align}
\noindent where the $^\circ$ superscript notation denotes a quantity that is annihilated by $\thorn$~\cite{kinnersley1969type}. Eq.~\eqref{eq:HeldCom} can be interpreted as $\thph$, $\edthh$, and $\edthhp$ containing no explicit $\rho$ dependency\footnote{Held integration is still limited to equations where only $\thorn$ acts on the unknown quantity, whether using Held operators or GHP operators; like \cref{eq:edthD}, $\edthh D[\rho] = E^\circ \rho^3$ cannot be solved for $D[\rho]$ using Held integration.}. 

To begin extracting the $\rho$ dependency in the spin coefficients in Kerr, Held used a relation for $\tau$ taken from Ref.~\cite{kinnersley1969type},
\begin{align}\label{eq:taucirc}
    \tau=\rho\bar{\rho}\tau^\circ,
\end{align}
with $\tau^\circ\circeq\{-1,-3\}$.
\noindent 
While Eq.~\eqref{eq:taucirc} was derived in the Kinnersley tetrad in Ref.~\cite{kinnersley1969type}, it holds in all Petrov type-D aligned tetrads~\cite{Hollands:2024iqp}.

To extract the $\rho$ dependency from the other spin coefficients, Held integrated various Ricci identities, Bianchi identities, and commutation relations~\cite{held1974formalism}, finding
\begin{align}\label{eq:taupcirc}
    \tau' &= -\rho^2\tab^\circ,\\
    \psi_2 &= \Psi^\circ \rho^3,\label{eq:psicirc} \\
    \rho' &= \rho^{\prime \circ} \bar \rho - \frac{1}{2} \Psi^\circ \rho^2 - \left(\edthp \tab^\circ + \frac{1}{2}\Psi^\circ\right) \rho \bar \rho - \tau^\circ \tab^\circ \rho^2 \bar \rho,\label{eq:rhopcirc}
\end{align}
which defines the Held scalars $\Psi^\circ\circeq\{-3,-3\}$ and $\rho^{\prime \circ}\circeq\{-2,-2\}$.

There is an additional Held scalar that can be derived from the spin coefficient $\rho$,
\begin{align}\label{eq:Omegacirc}
    \Omega^\circ:= \frac{1}{\rhob} -\frac{1}{\rho},
\end{align}
with $\Omega^\circ\circeq\{-1,-1\}$.

At this point, we will deviate from the conventional Held formalism notation. In the Kinnersley tetrad we can express the Held quantities in a coordinated form. In Boyer-Lindquist coordinates ($\{t,r,\theta,\phi\}$)~\cite{green2020teukolsky, toomani2021new},
\begin{align}
\tau^\circ &= -\frac{ia\sin[\theta]}{\sqrt{2}},\qquad \rho^{\prime\circ}=-1/2,\label{eq:Heldcoeffs}  \\
 \qquad \Omega^\circ &=-2ia\cos[\theta], \ \ \ \ \ \Psi^\circ = M.\label{eq:Heldcoeffs2}
\end{align}
Clearly, $\rho^{\prime\circ}$ and $\Psi^\circ$ are coordinate invariant. In the following, we replace $\rho^{\prime\circ}$ and $\Psi^\circ$ with $-1/2$ and $M$ respectively. Note, this relabelling does not have a well-defined spin and boost weight, but this does not affect the form of Held equations because $\th \Psi^\circ=\thph \Psi^\circ=\edthh \Psi^\circ= \edthhp \Psi^\circ=\th \rho^{\prime\circ}=\thph \rho^{\prime\circ}=\edthh \rho^{\prime\circ}= \edthhp \rho^{\prime\circ} =0$ (from the Ricci and Bianchi identities~\cite{held1974formalism}).


Held used the remaining Ricci and Bianchi identities (and commutation relations) to derive simplifications for the Held derivatives acting on the Kerr--Held scalars,
\begin{align}
\thph \rho&=-\frac{\rho^2}{2}-\frac{\rho^2M}{2}(\rho+\rhob)-\rho^3\rhb\tau^\circ, \label{eq:HeldSimps1} \\ \edthh \rho&=\rho^2\tau^\circ, \ \edthhp \rho=-\rho^2\taub^\circ, \label{eq:HeldSimps2} \\
     \thph \tau^\circ&=0, \ \edthh \tau^\circ=0, \ \edthhp \tau^\circ =\frac{1}{2}\bar{\Omega}^\circ, \label{eq:HeldSimps3} \\
        \thph \Omega^\circ&=0, \ \edthh \Omega^\circ=2\tau^\circ, \ \edthhp \Omega^\circ =-2\taub^\circ.\label{eq:HeldSimps4}
\end{align}
\noindent 

\subsubsection{Held derivatives and spin-raising/lowering operators}

A further convenience of the Held formalism is that $\edthh$ and $\edthhp$ relate to the spin-raising and lowering operators of the spin-weighted spherical harmonics. Next, we comment on how this relationship is not unique to the Held formalism and how the Kinnersley tetrad is one of a select class of tetrads where the GHP derivatives $\edth$ and $\edthp$ relate to the spin-raising and lowering operators.

In the Kinnersley tetrad, it is not challenging to extract the spin-raising and lowering operators from the GHP derivatives $\edth$ and $\edthp$ (Eq.~\eqref{eq:GHPderivatives}) when expressed in coordinate form. First, express the GHP derivatives $\edth$ and $\edthp$ in NP form using \cref{eq:GHPderivatives}. From \cref{eq:GHPderivatives}, we see the relevant NP quantities are $\delta$, $\beta$, $\alpha$, and their complex conjugates. In the Kinnersley tetrad and Boyer--Lindquist coordinates, \cref{eq:Kinnersley-l,eq:Kinnersley-n,eq:Kinnersley-m,eq:Kinnersley-mbar} give
\begin{align}
    \delta&=\frac{1}{\sqrt{2}\bar\zeta}\Big(  ia\sin[\theta]\partial_t + \partial_\theta + \frac{i}{\sin[\theta]}\partial_\phi  \Big), \label{eq:delta}\\
    \beta&=\frac{\cot[\theta]}{2\sqrt{2}\bar\zeta}, \ \ -\alpha=\beta'=\frac{\cot[\theta]}{2\sqrt{2}\zeta} -\frac{ia \sin[\theta]}{\sqrt{2}\zeta^2}.\label{eq:beta}
\end{align}
Noting the common factor of $\frac{1}{\sqrt{2}\bar\zeta}$ in \cref{eq:delta,eq:beta}, and using \cref{eq:GHPderivatives}, it is straightforward to show the GHP $\edth$ relates to the spin raising operator $\hat\edth$,
\begin{align}\label{eq:GHPEdth}
\edth & = \frac{1}{\sqrt{2}\bar\zeta} \hat\edth + \frac{ia\sin[\theta]\partial_{u}}{\sqrt{2}\bar\zeta}+\frac{iq\sin[\theta]}{\sqrt{2}\bar\zeta^2}. 
\end{align}

The Kinnersley tetrad is not the only tetrad where GHP $\edth$ and $\edthp$ relate to spin-raising and lowering operators. This can be seen by examining the class III tetrad rotations~\cite{chandrabook} (under which the direction of the vectors $l^a$ and $n^a$ are unchanged and, therefore, remain aligned with the principle null directions),
\begin{align}\label{eq:tetradtransform}
    l^a\rightarrow A^{-1}l^a,  \ n^a\rightarrow A n^a,  \ m^a\rightarrow e^{i\theta}m^a,  \ \bar m^a\rightarrow e^{-i\theta}\bar m^a,
\end{align}
where $A$ and $\theta$ are real functions. The relevant spin coefficients (see \cref{eq:GHPderivatives}) transform as~\cite{chandrabook},
\begin{align}
    \beta &\rightarrow e^{i\theta} \beta +\frac12 i e^{i\theta} \delta \theta - \frac12 A^{-1} e^{i\theta} \delta A, \\
    \beta' &\rightarrow e^{-i\theta} \beta' -\frac12 i e^{-i\theta} \bar\delta \theta + \frac12 A^{-1} e^{-i\theta} \bar\delta A,
\end{align}
Take a tetrad related to the Kinnersley tetrad where $\theta$ and $A$ satisfy $\delta \theta=\delta A=0$; then $\beta$, $\bar\beta'$, and $m^a$ are only re-scaled by a factor of $e^{i\theta}$, so the relation of the GHP derivatives $\edth$ and $\edthp$ to the spin raising and lowering operators remains trivial. The Hartle--Hawking tetrad~\cite{hawking1972energy, poisson2004absorption} is one such example, where $A=\frac{2(r^2+a^2)}{\Delta}$ and $\theta=0$. However, this condition does not hold for all tetrads, such as the Carter tetrad. To transform from the Kinnersley to the Carter tetrad $A=\sqrt{\frac{2\Sigma}{\Delta}}$ and $\theta=-i\ln\Big[\frac{\bar\zeta}{\sqrt{\Sigma}}\Big]$. Hence, 
\begin{align}\label{eq:Carterm}
    m^a_C = \frac{1}{\sqrt{2\Sigma}}\Big\{  ia\sin[\theta], 0 , 1, \frac{i}{\sin[\theta]}  \Big\}
\end{align}
and the relevant spin coefficients in the Carter tetrad are
\begin{align}\label{eq:CarterB}
    \beta_C=\beta'_C=-\frac{i}{\zeta}\frac{a+ir\cos[\theta]}{2\sin[\theta]\sqrt{2\Sigma}}.
\end{align}
Examining \cref{eq:Carterm,eq:CarterB}, in the Carter tetrad, the $\cot[\theta]$ terms in $\beta$ and $\beta'$ do not have the same coefficient as the $\theta$ derivative in $\delta$. Hence the $\cot[\theta]$ terms are not eliminated by the conversion to a spin-raising operator. Also, the $\frac{a}{\sin[\theta]}$ terms in \cref{eq:CarterB} will remain after expressing the $\theta$ derivatives in terms of raising and lowering operators. $\cot[\theta]$ and $\frac{1}{\sin[\theta]}$ are singular at the poles, and when expressed as a sum of spin-weighted spherical harmonics, the sum does not converge. Therefore, there are preferred tetrads for expressing Kerr quantities such that they are smooth on the 2-sphere: the Kinnersley tetrad and tetrads related to the Kinnersley tetrad by \cref{eq:tetradtransform} with $\delta \theta=\delta A=0$. 

While the Kinnersley tetrad in Boyer--Lindquist coordinates avoids singularities in $\edth$ and $\edthp$, we can see from \cref{eq:Kinnersley-l,eq:GHPderivatives} that $\thorn$ contains $\frac{1}{\Delta}$ terms. These terms are singular at the horizons as $\Delta=0$ for
\begin{align}\label{eq:rplus}
    r_\pm=M\pm\sqrt{M^2-a^2},
\end{align}
the outer and inner horizons, respectively. To avoid singularities at the horizon and make $\thorn$ and $\tilde{\th}'$ as simple as possible, we work in Kerr--Newman coordinates~\cite{green2020teukolsky}.
 The outgoing Kerr--Newman coordinates, $\big\{u,r,\theta, \phi_* \big\}$, are related to Boyer--Lindquist coordinates as follows,~\cite{green2020teukolsky, KerrNewmanCoords}
\begin{align}
    u&=t-r_*=t-r-\frac{r_+^2+a^2}{r_+-r_-} \ln\Big[ \frac{r-r_+}{r_+}\Big] \notag \\ & \ \ \ \ \ \ \ \ \ \ \ \ \ \ \ \ \ \ \  \ \ \ +\frac{r_-^2+a^2}{r_+-r_-} \ln\Big[ \frac{r-r_-}{r_-} \Big], \\
    \phi_* &= \phi -\frac{a}{r_+-r_-} \ln\Big[ \frac{r-r_+}{r-r_-} \Big],
\end{align}
 In these coordinates~\cite{green2020teukolsky, toomani2021new},
\begin{align}
    l^a&=\Big\{  0, 1, 0 , 0 \Big\},\label{eq:KN-l}  \\
    n^a&=\frac{1}{\Sigma}\Big\{   r^2+a^2, -\frac{\Delta}{2}, 0 , a  \Big\},\label{eq:KN-n} \\
    m^a&=\frac{1}{\sqrt{2}\bar\zeta}\Big\{   ia\sin[\theta], 0, 1 , \frac{i}{\sin[\theta]}  \Big\}, \label{eq:KN-m}\\
    \bar m^a&=\frac{1}{\sqrt{2}\zeta}\Big\{   -ia\sin[\theta], 0, 1 , \frac{-i}{\sin[\theta]}  \Big\};    \label{eq:KN-mbar} 
\end{align}
Using Held's definitions~\cite{held1974formalism}
in the Kinnersley tetrad, it has been shown in Ref.~\cite{Hollands:2024iqp},
\begin{align}\label{eq:thorn}
    \thorn&=\partial_r ,
\end{align}
and
\begin{align}\label{eq:thphrho}
    \tilde{\th}'  & =\partial_{u}-\frac{\Delta}{2\Sigma}\partial_{r} \notag \\
    &=\partial_{u}-\frac{\Delta}{2}\rho\rhob{} \partial_{r}.
\end{align}

The null vectors $m^a$ and $\bar m^a$ in \cref{eq:KN-m,eq:KN-mbar} take the same form as in Boyer--Lindquist coordinates (as do the spin-coefficients as they are scalars which depend on only $r$ and $\theta$). The Held derivatives $\edthh$ and $\edthhp$ then relate similarly to the spin raising and lowering operators,
\begin{align}\label{eq:HeldEdth}
\edthh & = \frac{-1}{\sqrt{2}} \left(\partial_{\theta} + i \csc [\theta]\partial_{\phi_*} + ia \sin [\theta]\partial_{u} - \frac12 (p-q) \cot [\theta] \right)  \notag \\
&=\frac{-1}{\sqrt{2}} \hat\edth - \frac{ia\sin[\theta]\partial_{u}}{\sqrt{2}}, \\
\edthhp   &= \frac{-1}{\sqrt{2}} \left(\partial_{\theta} - i \csc [\theta]\partial_{\phi_*} - ia \sin [\theta]\partial_{u} + \frac12 (p-q) \cot [\theta] \right)  \notag \\
&=\frac{-1}{\sqrt{2}} \hat\edthp+\frac{ia\sin[\theta]\partial_{u}}{\sqrt{2}},\label{eq:HeldEdthp}
\end{align}
with the subtlety that the spin raising and lowering operators now act on ${}_s Y_{\l\m}[\theta,\phi_*]$ rather than ${}_s Y_{\l\m}[\theta,\phi]$. The $\partial_{u}$ derivatives are trivial as the background is time-independent and we assume the perturbations have a time dependency proportional to $e^{-i\omega t}$ (converting the time coordinate dependence is straightforward, $e^{-i\omega t}=e^{-i\omega (u+r_*)}$). \cref{eq:HeldEdth,eq:HeldEdthp} are similar to Chandrasekhar's operators  $\chand{-s}{\omega}^\dagger, \chand{+s}{\omega}$ up to a trivial normalization~\cite{chandrabook}.

\subsubsection{Converting to coordinates and modedecomposition}\label{sec:coords}

To convert expressions in the Held formalism into Kerr--Newmann coordinates ($\{u,r,\theta,\phi_*\}$), we first express the derivatives $\th$ and $\thph$ using \cref{eq:thorn,eq:thphrho}. The resulting expression presents all the $\rho$ and $\rhob$ dependence explicitly\footnote{One may encounter $\partial_i \rho$ For $i=\{u,r,\phi_*\}$ terms, it is trivial to show $\partial_r \rho =\rho^2$, $\partial_u \rho =\partial_{\phi_*} \rho=0$, and similarly for $\rhob$.}. Next, we deal with the remaining angular dependence and $\edthh $ and $\edthhp $ operators.


%
%
We assume the inputs ($A$) are expressed as a convergent sum of spin-weighted spherical harmonics with coefficients that can depend on $r$, $\rho$, and $\rhob$:
\begin{align}
    A=\sum_{\l= |s|}^\infty \sum_{\m=-\l}^{\l} A_{\l\m}[r,\rho,\rhob] \ {}_s Y_{\l\m}[\theta,\phi_*] \ e^{i\omega u},\label{eq:Adecomp}
\end{align}
where $A_{\l\m}[r, \rho,\rhob]$ are polynomials with positive and finite powers of $\rho$ and $\rhob$; that is, $A_{\l\m}[r, \rho,\rhob]= \rho A^{1,0}_{\l\m}[r]+\rhob A^{0,1}_{\l\m}[r] +\rho\rhob A^{1,1}_{\l\m}[r]+ \ldots +\rho^i\rhob^j A^{i,j}_{\l\m}[r]$ for some finite $i$ and $j$. Note that $\phi_*$ derivatives, like time derivatives, are trivial as ${}_s Y_{\l\m}\sim e^{i\m\phi_*}$; that is,
\begin{align}
    \partial_{\phi_*} A = i\m A.
\end{align}

While we can use Eq.~\cref{eq:Heldcoeffs,eq:Heldcoeffs2} to express the Held scalars in coordinate form, for a mode decomposition, it is more useful to express them in terms of spin-weighted spherical harmonics\footnote{\cref{eq:Heldcoeffs,eq:Heldcoeffs2} are identical in Kerr--Newman coordinates as in BL coordinates as they are scalars which only depend on $r$ and $\theta$.},

\begin{align}
    \tau^\circ=-\sqrt{\frac{4\pi}{3}}a\ {}_{1} Y_{10},& \ \ \taub^\circ=-\sqrt{\frac{4\pi}{3}}a\ {}_{-1} Y_{10}, \notag \\ \Omega^\circ =-4i&\sqrt{\frac{\pi}{3}}a\ {}_{0} Y_{10};\label{eq:HeldscalarY}
\end{align}
%
where we have used
\begin{align}\label{eq:reexpandZ}
    Z[\theta] =\sum_{\l=s,0}^\infty  c^{\l,0,s}{}_{s}Y_{\l,0},
\end{align}
with
\begin{align}\label{eq:cintegral}
    c^{\l,0,s}=\oint {}_{s}\bar{Y}_{\l,0}Z[\theta].
\end{align}

The resulting source expression will have products of many spin-weighted spherical harmonics. To combine the spin-weighted spherical harmonics, we use \cref{eq:reexpandY}. The resulting expression for the source will be of the form
\begin{align}\label{eq:sourceYnorho}
    f[t,r,\theta,\phi]=  \sum_{\l=|s|}^\infty \sum_{\m=-\l}^{\l} \tilde f_{\l\m}[r, \rho,\rhob]e^{-i\omega t} {}_s Y_{\l\m}[\theta,\phi_*].
\end{align}
Where $\tilde f_{\l\m}[r, \rho,\rhob]$ is a polynomial with positive and finite powers of $\rho$ and $\rhob$; that is, $\tilde f_{\l\m}[r, \rho,\rhob]= \rho \tilde f^{1,0}_{\l\m}[r]+\rhob \tilde f^{0,1}_{\l\m}[r] +\rho\rhob \tilde f^{1,1}_{\l\m}[r]+ \ldots +\rho^i\rhob^j \tilde f^{i,j}_{\l\m}[r]$ for some finite $i$ and $j$. Next, we must express $\rho$ and $\rhob$ in terms of spin-weighted spherical harmonics.

\subsection{$\rho$ expansion}\label{sec:rhoexpansions}

So far, our method does not completely express quantities separably as factors where $\rho$ and $\rhob$ are present; see Eq.~\eqref{eq:rho}. The mixing between the radial and polar angle coordinate in the denominator of Eq.~\eqref{eq:rho} makes separating the $r$ and $\theta$ dependency non-trivial. In this subsection, we show the $r$ and $\theta$ dependency in $\rho^i\rhob^k$ (for any positive integer $i$ and $k$) can be separated using a Fourier expansion\footnote{Alternatively, one can separate the dependency using a two-point Taylor series~\cite{estes1966two, locke_2020, o2022conservative}; however, we found the Fourier series approximation more accurate.}. 

As Eq.~\eqref{eq:rho} is a trigonometric function, a Fourier expansion seems like a natural approach. Eq.~\eqref{eq:rho} is periodic in $\theta$ ($\rho[r,\theta]=\rho[r,\theta+2n\pi]$ for any $r$ and $\theta$, where $n$ is an integer). Also, Eq.~\eqref{eq:rho} is complex and non-singular for $r\geq r_-$ (where $r_-$ is the radius of the inner horizon). To assess the real and imaginary behaviour of Eq.~\eqref{eq:rho}, one can write
\begin{align}\label{eq:rhoReIm}
    \rho=\frac{-(r+ia\cos[\theta])}{\Sigma},
\end{align}
as 
\begin{align}
    \frac{1}{\Sigma}=\rho\rhob=\frac{1}{(r^2+a^2 \cos^2[\theta])}.
\end{align}
All factors of $\rho^i\rhob^k$ for $i\neq k$ can also be written in a manner where the denominator is real: for $i\geq k$,
\begin{align}\label{eq:rhoReIm1}
    \rho^i\rhob^k=\frac{(-(r+ia\cos[\theta]))^{i-k}}{\Sigma^i},
\end{align}
and for $k\leq i$,
\begin{align}\label{eq:rhoReIm2}
    \rho^i\rhob^k=\frac{(-(r-ia\cos[\theta]))^{k-i}}{\Sigma^k}.
\end{align}
This form is useful as the Fourier expansion of the numerators in Eqs.~\eqref{eq:rhoReIm},~\eqref{eq:rhoReIm1} and~\eqref{eq:rhoReIm2} are trivial. Hence, only the Fourier expansion of the denominator, which is real, is required. 

A common denominator factor of $\Sigma$ is desirable to minimise the number of Fourier expansions required. Using Eq.~\eqref{eq:rhoReIm}, we can express Eq.~\eqref{eq:sourceYnorho} as
\begin{align}\label{eq:sourceYrho}
    f[t,r,\theta,\phi]= \frac{1}{\Sigma^i} \sum_{\l=|s|}^\infty \sum_{\m=-\l}^{\l}  \breve f_{\l\m}[r]e^{-i\omega t} {}_s Y_{\l\m}[\theta,\phi_*].
\end{align}
for some $i$, which has collected all of the factors of $\rho$ and $\rhob$ outside of the sum in the form of $\frac{1}{\Sigma^i}$. Therefore, we require only one Fourier expansion, that of $\frac{1}{\Sigma^i}$. For example, take the function 
\begin{align}
   f= A^\circ\rho+B^\circ \rho^3\rhob^2+C^\circ \rhob^4.
\end{align}
To extract the same $\rho\rhob$ dependency, we note that the highest power of either $\rho$ or $\rhob$ is $\rhob^4$. Hence, we write this function as,
\begin{align}
    f&=\rho^4\rhob^4(\frac{A^\circ}{\rho^3\rhob^4}+\frac{B^\circ}{ \rho\rhob^2}+\frac{C^\circ}{\rho^4}).
\end{align}
Using Eq.~\eqref{eq:rho} to replace the inverse powers of $\rho$ and $\rhob$ gives
\begin{align}
    &f=\frac{1}{\Sigma^4}(-(r-ia\cos[\theta])^3(r+ia\cos[\theta])^4A^\circ \notag \\
    &-(r-ia\cos[\theta])(r+ia\cos[\theta])^2B^\circ+(r-ia\cos[\theta])^4C^\circ).
\end{align}
We can use $\cos[\theta]=\sqrt{\frac{4\pi}{3}}\ {}_0 Y_{10}$ and Eqs.~\eqref{eq:reexpandY} and \eqref{coupling} to combine the angular dependence with that in $A^\circ$, $B^\circ$, and $C^\circ$, allowing us to express $f$ in the form of Eq.\eqref{eq:sourceYrho}. 


\subsubsection{Fourier series}

\begin{figure}[t]

{\includegraphics{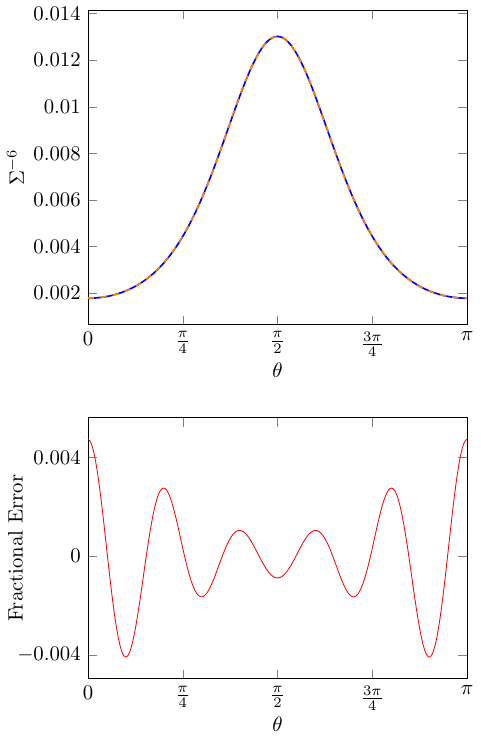}}

\caption{\label{fig:Fourier-accuracy} Top: the function $\Sigma^{-6}$ is plotted in blue (for $a=0.9$ and $r=r_+\approx1.436M$), and the Fourier expansion (for $k=8$, see \cref{eq:FourierSeries}) is over-plotted as a dashed yellow line. Bottom: the fractional disagreement between the two plots is shown. There is less than $0.5\%$ error, and the error decreases for increasing radius (see Fig.~\ref{fig:Precision-accuracy9}).}
\end{figure}

\begin{figure}[t]

{\includegraphics{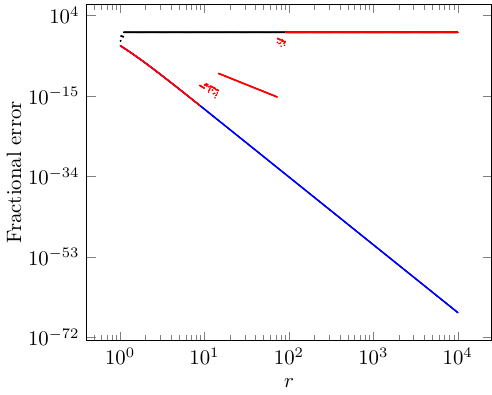}}

\caption{\label{fig:Precision-accuracy9} The fractional error, $|\frac{\Sigma^{-6}}{F[\Sigma^{-6},k=12]}-1|$, for different digits of precision used (black line is 8, red line is 32, and blue line is 128 digits of precision), for $a=0.9$ and $\theta=\frac{\pi}{4}$. The plots show a clear dependence of the error on the number of digits of precision used for large $r$. 
This problem does not occur for $k<2i$ when approximating $\Sigma^{-i}$.}
\end{figure}

A Fourier series expresses periodic functions using the orthogonality of the trigonometric functions $\sin[n\theta]$ and $\cos[n\theta]$. Truncated to order $k$, over an interval of $2\pi$, the Fourier series of a function $A[\theta]$ can be expressed as
\begin{align}\label{eq:FourierSeries}
    F[A[\theta],k]=\frac{a_0}{2} +\sum_{n=1}^{k} a_n\cos[n\theta]+\sum_{n=1}^{k} b_n\sin[n\theta],
\end{align}
where
\begin{align}
    a_0&=\frac{1}{\pi}\int_{-\pi}^{\pi} A[\theta] d\theta,  \\
    a_n&=\frac{1}{\pi}\int_{-\pi}^{\pi} A[\theta]\cos[n\theta] d\theta,  \\
    b_n&=\frac{1}{\pi}\int_{-\pi}^{\pi} A[\theta]\sin[n\theta] d\theta .
\end{align}

The quantity $\frac{1}{\Sigma^i}$ is an even periodic function over $[-\pi,\pi]$; hence, $b_n=0$. Also, as $\Sigma^i$ is a function of only even powers of $\cos[\theta]$, $a_{2n+1}=0$ for any positive integer $n$. Note that $\frac{1}{\Sigma^i}$ is not singular in the domain $r\in [r_{+},\infty)$ so we expect our Fourier series to converge.

We now assess the accuracy of a truncated Fourier series. For example, take the expansion of $\Sigma^{-6}$ to order $k=8$,
\begin{align}\label{eq:Fouriersigma4}
    F[&\Sigma^{-6},8]=\frac{1}{256 r^{11} (a^2 + r^2)^{11/2}}\Big( (a^2 + 2 r^2) (63 a^8 \notag \\
    &+ 224 a^6 r^2 + 352 a^4 r^4 + 256 a^2 r^6 + 128 r^8) - 6 a^2 (21 a^8 \notag \\
    &+ 112 a^6 r^2 + 240 a^4 r^4 + 256 a^2 r^6 + 128 r^8) \cos[2\theta] \notag \\
    &+ 42 a^4 (a^2 + 2 r^2) (3 a^4 + 8 a^2 r^2 + 8 r^4) \cos[4\theta] \notag \\
    &- 14 a^6 (9 a^4 + 32 a^2 r^2 + 32 r^4) \cos[6\theta]\notag \\
    & + 126 a^8 (a^2 + 2 r^2) \cos[8\theta]\Big),
\end{align}
calculated using \textit{Mathematica}~\cite{Mathematica}. Equation~\eqref{eq:Fouriersigma4} accurately approximates $\Sigma^{-6}$, as can be seen in Fig.~\ref{fig:Fourier-accuracy}. For $a=0.9$, $r=1.43M$, and $k=8$, the error is $<0.5\%$. For larger $r$, the approximation becomes more accurate (similarly to the blue line in Fig.~\ref{fig:Precision-accuracy9}). The most inaccurate region is for small $r$ and near the poles. For $a=0.9$, the outer black hole horizon is $r_+\approx 1.436M$ (\cref{eq:rplus}). For higher values of $a$, the truncated Fourier series is less accurate near the horizon\footnote{This is due to the horizon radius approaching $r=1M$ as $a\rightarrow 1$; therefore the $-ia\cos[\theta]$ part in the denominator of \cref{eq:rho} will have a similar magnitude to $r$, $\frac{1}{\Sigma}$ observes similar behaviour. That is, the angular dependence becomes more significant. Hence, an accurate approximation requires truncating the series at a higher order in $k$ in \cref{eq:FourierSeries}.}. Considering the accuracy requirements of LISA and current estimates on the spins of supermassive black holes, the expansion used in Fig.~\ref{fig:Fourier-accuracy} is probably sufficient.

To achieve arbitrarily high accuracy, one can increase the number of terms in the Fourier expansion, $k$. However, when attempting to approximate $\Sigma^n$ with a Fourier series with $k\geq 2n$, errors can occur for large $r$ if sufficiently high precision is not used, as shown in Fig.~\ref{fig:Precision-accuracy9}. We examine the simplest example, $\Sigma^{-1}$, to understand why large errors are encountered when using low precision. The Fourier series of $\Sigma^{-1}$ with $k=2$ is
\begin{align}
    F[\Sigma^{-1},2]= \frac{ a^2 -2\big( a^2 
+2r(r- \sqrt{a^2+r^2}) \big)\cos[2\theta]   }{ra^2\sqrt{a^2+r^2}}
\end{align}
Naively, the Fourier series appears to scale as $\sim r^0$ for large $r$, whereas $\frac{1}{\Sigma}\sim r^{-2}$. However, $r- \sqrt{a^2+r^2}\rightarrow 0$ for large $r$, so this contribution is suppressed, so the Fourier series scales as $r^{-2}$, as expected. However, resolving the suppression of such terms numerically requires high precision. With insufficient precision, an error $\propto r^2$ is incurred. 


In practice, using such high precision will slow calculations down. 
Nonetheless, for EMRI models for LISA data analysis, the accuracy requirements are likely sufficiently modest that Fourier expansions with $k<2i$ can be used; that is, high precision evaluation will likely not be required. 

\subsection{Combining the spin-weighted spherical harmonics}

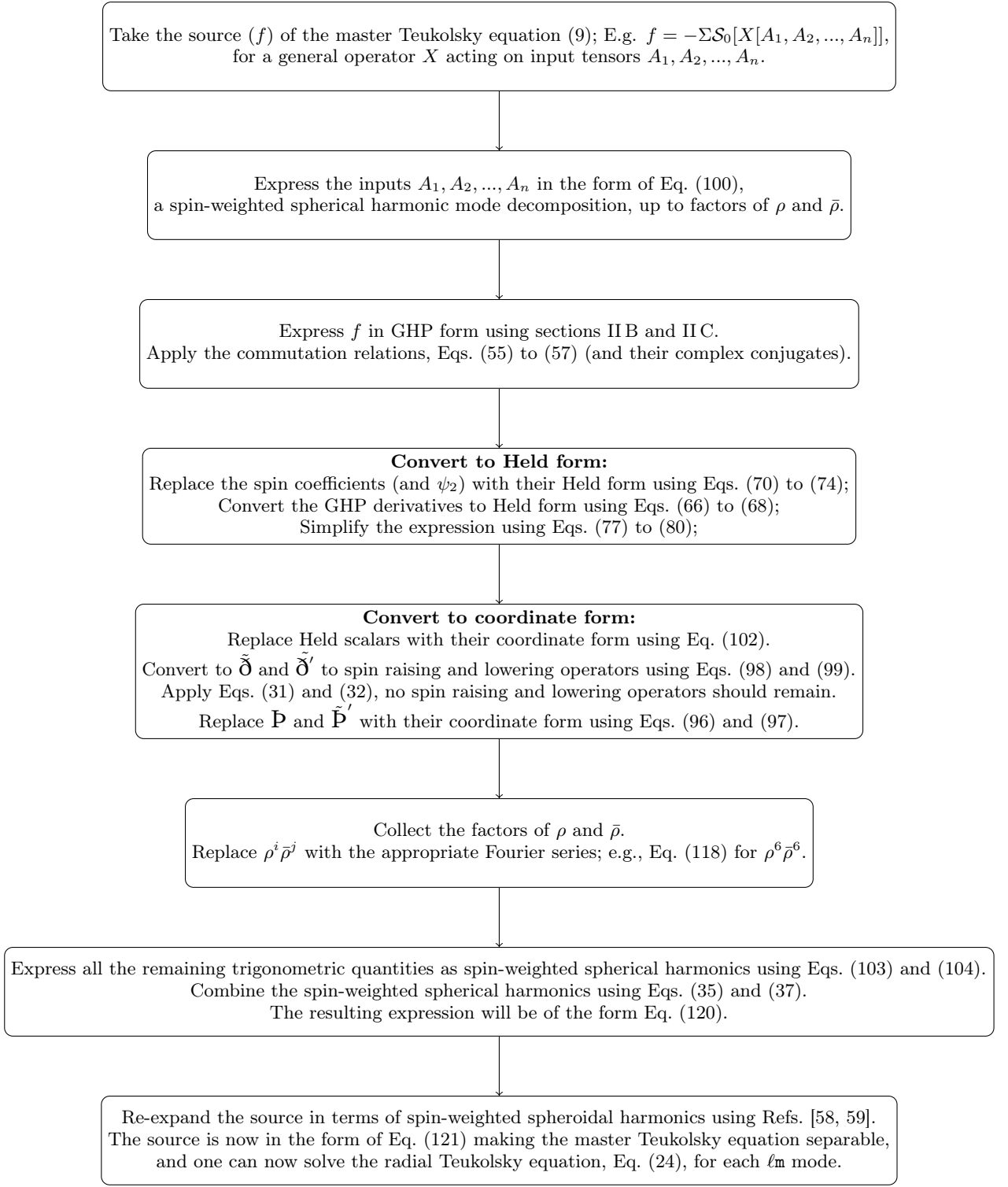
\begin{figure*}
    \centering

\begin{tikzpicture}[node distance=1cm, every node/.style={draw, rectangle, align=center, rounded corners, minimum width=8cm, minimum height=1.5cm}]
\node (box1) {Take the source ($f$) of the master Teukolsky equation~\eqref{eq:masterTeuk};
E.g. $f=-\Sigma \mathcal{S}_0[ X[A_1,A_2,...,A_n]]$,\\ for a general operator $X$ acting on input tensors $A_1,A_2,...,A_n$.};

\node (box2) [below=of box1] {Express the inputs $A_1,A_2,...,A_n$ in the form of \cref{eq:Adecomp},\\ a spin-weighted spherical harmonic mode decomposition, up to factors of $\rho$ and $\rhob$.};

\node (box3) [below=of box2] {Express $f$ in GHP form using sections~\ref{sec:NP}~and~\ref{subsec:GHP}.\\ Apply the commutation relations, \cref{eq:GHPCommutation1,eq:GHPCommutation2,eq:GHPCommutation3} (and their complex conjugates).};

\node (box4) [below=of box3] {\textbf{Convert to Held form:}\\

    Replace the spin coefficients (and $\psi_2$) with their Held form using \cref{eq:taucirc,eq:taupcirc,eq:psicirc,eq:rhopcirc,eq:Omegacirc};\\
    Convert the GHP derivatives to Held form using \cref{,eq:thph,eq:edthh,eq:edthph};\\
    Simplify the expression using \cref{eq:HeldSimps1,eq:HeldSimps2,eq:HeldSimps3,eq:HeldSimps4};

};

\node (box5) [below=of box4] {\textbf{Convert to coordinate form:}\\
    Replace Held scalars with their coordinate form using \cref{eq:HeldscalarY}.\\
    Convert to $\edthh$ and $\edthhp$ to spin raising and lowering operators using \cref{eq:HeldEdth,eq:HeldEdthp}. \\
    Apply \cref{eq:ethswshdef,eq:ethprimeswshdef}, no spin raising and lowering operators should remain.\\
    Replace $\th$ and $\thph$ with their coordinate form using \cref{eq:thorn,eq:thphrho}.

};

\node (box6) [below=of box5] {Collect the factors of $\rho$ and $\rhob$. \\
Replace $\rho^i\rhob^j$ with the appropriate Fourier series; e.g., \cref{eq:Fouriersigma4} for $\rho^6\rhob^6$.
};

\node (box7) [below=of box6] {Express all the remaining trigonometric quantities as spin-weighted spherical harmonics using \cref{eq:reexpandZ,eq:cintegral}. \\
Combine the spin-weighted spherical harmonics using \cref{eq:reexpandY,coupling}.
\\ The resulting expression will be of the form \cref{eq:sourceY2}.
};

\node (box8) [below=of box7] {Re-expand the source in terms of spin-weighted spheroidal harmonics using Refs.~\cite{hughes2000evolution,BHPToolkit}.\\
The source is now in the form of \cref{eq:sourceS2} making the master Teukolsky equation separable, \\
and one can now solve the radial Teukolsky equation, \cref{eq:seperableTeuk}, for each $\l\m$ mode. 
};

\draw[->] (box1) -- (box2);
\draw[->] (box2) -- (box3);
\draw[->] (box3) -- (box4);
\draw[->] (box4) -- (box5);
\draw[->] (box5) -- (box6);
\draw[->] (box6) -- (box7);
\draw[->] (box7) -- (box8);

\end{tikzpicture}

    \caption{\label{fig:Flow-chart} Summary of our method for separating the source of the Teukolsky equation.}
    \label{fig:enter-label}
\end{figure*}

Finally, we can replace $\rho^{i }\rhob^{i}=\Sigma^{-i}$ factor in Eq.~\eqref{eq:sourceYrho} with the Fourier expansion truncated to some finite order. To convert the trigonometric quantities in the Fourier series with spin-weighted spherical harmonics, we can use \cref{eq:reexpandZ,eq:cintegral}. By again using Eqs.~\eqref{eq:reexpandY} and~\eqref{coupling}, we can combine the spin-weighted spherical harmonics to express the source as
\begin{align}\label{eq:sourceY2}
    f[t,r,\theta,\phi]=  \sum_{\l=|s|}^\infty \sum_{\m=-\l}^{\l}  \hat f_{\l\m}[r]e^{-i\omega t} {}_s Y_{\l\m}[\theta,\phi_*].
\end{align}
To solve the separable master Teukolsky equation, \cref{eq:seperableTeuk}, we need to express the source as a spin-weighted spheroidal expansion rather than a spin-weighted spherical harmonic expansion. Using the inversion of the spin-weighted spherical harmonic expansion of the spin-weighted spheroidal harmonics~\cite{hughes2000evolution, BHPToolkit}, one can re-expand \cref{eq:sourceY2} in terms of spin-weighted spheroidal harmonics,
\begin{align}\label{eq:sourceS2}
    f[t,r,\theta,\phi]=  \sum_{\l=|s|}^\infty \sum_{\m=-\l}^{\l}   f_{\l\m}[r]e^{-i\omega t} {}_s S_{\l\m}[\theta,\phi_*].
\end{align}
This completes our formalism as we have separated the source into the spin-weighted spheroidal harmonic modes of the Teukolsky equation. The radial Teukolsky equation~\eqref{eq:seperableTeuk} can now be solved with $f_{\l\m}[r]$ in \eqref{eq:sourceS2}. Our formalism is summarised in Fig.~\ref{fig:Flow-chart}.

\section{Example: First-order Teukolsky equation with an extended source}\label{sec:IdealGasExample}

In this section, we present a toy-model application of our method to show that it provides a high-accuracy expression of the source of the Teukolsky equation. The accompanying \textit{Mathematica} notebook~\cite{Teukolsky-source-decomposer-notebook} provides the implementation of this example for reference. This notebook was built upon other publicly available notebooks~\cite{npnotebook, 2nd-order-notebook}.

While the initial motivation for our method is separating the source of the second-order Teukolsky equation, it also applies to the more straightforward case of the first-order Teukolsky equation. For our toy model, we will apply our formalism to an extended source of the first-order Teukolsky equation~\cref{eq:teuk}. This is a sufficient example for applying our formalism because the second-order Teukolsky equation (\cref{eq:reducedTeuk0,eq:reducedTeuk4}, or the Campanelli--Lousto form~\cite{l-c}) are similar to the first order equation: the sources involve operators which depend on background Kerr quantities and the inputs can be expanded in terms of spin-weighted spherical harmonics. In upcoming papers, we will apply our formalism to the second-order self-force and quasi-normal mode problems.

The source of the spin $-2$ first-order Teukolsky equation is
\begin{align}
    f[t&,r,\theta,\phi]=\S_4[T_{ab}]= \notag \\
    &\frac12(\edth^\prime - \bar \tau -4\tau^\prime)  \Big(  (\thorn^\prime - 2\bar \rho^\prime) T_{n\bar m} - (\edth^\prime -\bar\tau)T_{nn} \Big) \notag \\
    & + \frac12(\thorn^\prime - \bar \rho^\prime -4\rho^\prime)  \Big(  (\edth^\prime - 2\bar \tau) T_{n\bar m} - (\thorn^\prime -\bar\rho^\prime)T_{nn} \Big) .\label{eq:S4}
\end{align}
For our example, we take \cref{eq:S4} as our source and will express it in the form of \cref{eq:sourceY2}. Following Fig.~\ref{fig:Flow-chart}, we will expand the stress-energy tensor in terms of spin-weighted spherical harmonics, similarly to \cref{eq:Adecomp}. We also express \cref{eq:S4} in the Held formalism as outlined in Sec.~\ref{sec:Held-Formalism}; please see the accompanying \textit{Mathematica} notebook~\cite{Teukolsky-source-decomposer-notebook} for the step-by-step application and the resulting formulas.

We require a stress-energy tensor with an extended radial profile. The stress-energy tensor of a perfect fluid is
\begin{align}
    T^{ab} = (\varrho + p )u^a u^b +pg^{ab}_{(0)},\label{eq:Tab}
\end{align}
where $\varrho$ is the density, $p$ is the pressure, $u^a$ is the four-velocity. To produce a simple example for applying our method, we assume $p=0$ and
\begin{align}
    \varrho= P e^{-\frac{\Lambda}{r}},
\end{align}
where $P$ and $\Lambda$ are constants. Similarly, we assume the fluid is stationary in the rest-frame of the black hole,
\begin{align}
    u^a=\Big\{1,0,0,0\Big\}.\label{eq:u}
\end{align}
Using \cref{eq:Tab,eq:u,eq:Kinnersley-n,eq:Kinnersley-mbar,eq:Kerr-metric}, 
\begin{align}
    T_{nn}&=\frac{\varrho\Delta^2}{4\Sigma^2}=\frac{\varrho\Delta^2}{4} \rho\rhob, \label{eq:Tnn}\\
    T_{n\bar m}&= \frac{-\varrho\Delta ia\sin[\theta]}{2\sqrt{2}\Sigma\zeta} =   \frac{\varrho\Delta ia\sin[\theta]}{2\sqrt{2}}  \rho^2\rhob, \label{eq:Tnmb} \\
    T_{\bar m\bar m} &= \frac{-\varrho a^2\sin^2[\theta]}{2\zeta^2}= \frac{-\varrho a^2\sin^2[\theta]}{2} \rho^2,\label{eq:Tmbmb}
\end{align}
where we have explicitly expressed the $\rho$ and $\rhob$ content in \cref{eq:Tnn,eq:Tnmb,eq:Tmbmb}. Using \cref{eq:reexpandZ,eq:cintegral}, we can decompose \cref{eq:Tnn,eq:Tnmb,eq:Tmbmb} into spin-weighted spherical harmonics,
\begin{align}
    T_{nn}&=\rho\rhob\sqrt{\pi}\frac{\varrho\Delta^2}{2} {}_0Y_{00}, \label{eq:TnnY}\\
    T_{n\bar m}& =   \rho^2\rhob \Big(-i\sqrt{\frac{\pi}{3}}\varrho a \Delta \Big)  {}_{-1}Y_{10}, \label{eq:TnmbY} \\
    T_{\bar m\bar m} &= \rho^2\Big(-2\sqrt{\frac{2\pi}{15}}\frac{\varrho a^2}{2}\Big) {}_{-2}Y_{20}.\label{eq:TmbmbY}
\end{align}
In general applications of our formalism, the components of the stress-energy tensor will be expressed as a sum of spherical harmonic modes, in the form of \cref{eq:Adecomp}. For our toy-model example, the simplicity of our stress-energy tensor has resulted in the mode decomposition being that of a single mode for each component (\cref{eq:TnnY,eq:TnmbY,eq:TmbmbY}).

Inserting \cref{eq:TnnY,eq:TnmbY,eq:TmbmbY} into the Held formalism version of \cref{eq:S4} allows us to extract all the $\rho$ and $\rhob$ factors, expressing the source in the form of~\cref{eq:sourceYnorho}. The highest factor is $\rho^6\rhob^4$. We can pull out a common factor of $\rho^6\rhob^4$ using \cref{eq:rho} and combine the spin-weighted spherical harmonics using  \cref{eq:reexpandY,coupling} to express
\begin{align}\label{eq:sourceYrhoIdealGas}
    f[t,r,\theta,\phi]= \rho^6\rhob^4 \sum_{\l=|s|}^\infty \sum_{\m=-\l}^{\l}  \breve f_{\l\m}[r] {}_{-2} Y_{\l\m}[\theta,\phi_*].
\end{align}
Note there is no $e^{-i\omega t}$ dependence, and the spin-weighted spheroidal harmonics are equivalent to spin-weighted spherical harmonics because the source is stationary ($\omega=0$). 

To complete our decomposition method requires expressing $\rho^6\rhob^4$ in a separable form and combining it into the mode summation. To do so, we can use the Fourier expansion (with $k=8$) for $\Sigma^{-6}$ in \cref{eq:Fouriersigma4}, as $\rho^6\rhob^4=\Sigma^{-6} (-r-ia\cos[\theta])^2$. The resulting expression, after again applying \cref{eq:reexpandY,coupling}, is of the form
\begin{align}\label{eq:sourceIdealGasDecomp}
    f[t,r,\theta,\phi]=  \sum_{\l=|s|}^\infty \sum_{\m=-\l}^{\l}  \hat f_{\l\m}[r] {}_{-2} Y_{\l\m}[\theta,\phi_*].
\end{align}
The explicit form of $\hat f_{\l\m}[r]$ can be found in the accompanying \textit{Mathematica} notebook.

\begin{figure}[t]
{\includegraphics{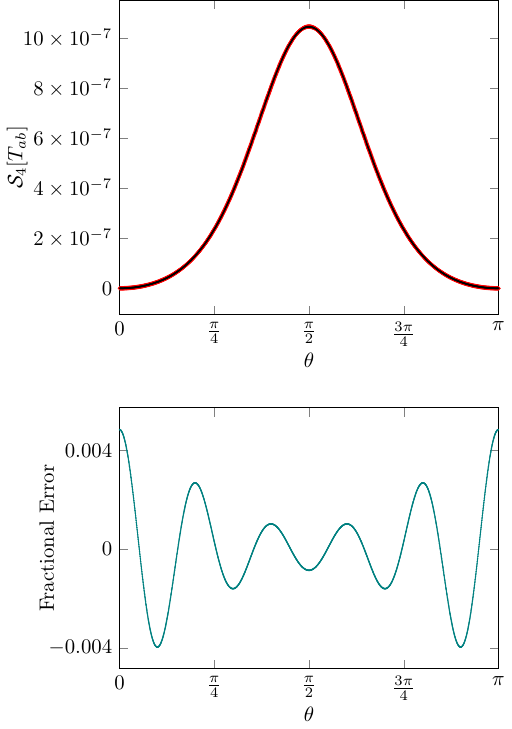}}

\caption{\label{fig:Ideal-gas-accuracy} Top: the source of the Teukolsky equation for a stationary ideal gas cloud is plotted in black (calculated using our decomposition method) and red (calculated explicitly) for $a=0.9$, $\Lambda=1$, $P=1$, and $r=r_+\approx1.436M$. The lines overlay each other, showing our decomposition method is accurate.  Bottom: the fractional disagreement between the two plots is shown; note the error is the same as Fig.~\ref{fig:Fourier-accuracy}, as expected.}
\end{figure}

In order to validate our expression for \cref{eq:sourceYrhoIdealGas}, we compare it to the four-dimensional expression for the source, \cref{eq:S4}. It is straightforward to calculate
\begin{align}\label{eq:4DSource}
    &f[t,r,\theta,\phi]= \notag \\
    &\frac{P a^2 e^{-\frac{r}{\Lambda}} \pi (2\Lambda + i a \cos{[\theta]} + r) (a^2 + r(-2M + r))^2 \sin^2{[\theta]}}{2\Lambda^2 (a \cos{[\theta]} + i r)^4 (i a \cos{[\theta]} + r)^3},
\end{align}
due to the simplicity of the stress-energy tensor using \cref{eq:GHPderivatives,eq:Kinnersley-l,eq:Kinnersley-n,eq:Kinnersley-m,eq:Kinnersley-mbar} (see Ref.~\cite{wardellpoundreview2021} for expressions of the spin-coefficients in the Kinnersley tetrad). We compare our four-dimensional source to the sum in \cref{eq:sourceYrhoIdealGas} in Fig.~\ref{fig:Ideal-gas-accuracy} near the horizon ($r=r_+\approx1.436M$). In the top plot, the two data sets overlap each other. The bottom plot shows that the error is less than $0.5\%$ and takes the same form as the error in Fig.~\ref{fig:Fourier-accuracy}, as expected.

One could decompose \cref{eq:4DSource} into spin-weighted spherical harmonics using \cref{eq:Soint} (with spin-weighted spherical harmonics instead of spin-weighted spheroidal harmonics). Solving \cref{eq:Soint} requires computing an integral for each $\l\m$ mode either analytically or numerically. For our toy-model example, the integrals are very slow to compute analytically; for second-order sources, which are much more complicated functions of $r$ and $\theta$ (see Ref.~\cite{Spiers:2023Coupling} and the \textit{PerturbationEquations} package in Ref.~\cite{BHPToolkit} for examples in Schwarzschild), analytical integration would be impractical. Calculating the integrals numerically requires computing an integral at each radial point on a grid, which is inefficient. Our formalism decomposes the source into modes without calculating integrals, so we expect it to provide efficiency savings for second-order calculations where the equations for the source are very long.

\begin{figure}[t]

{\includegraphics{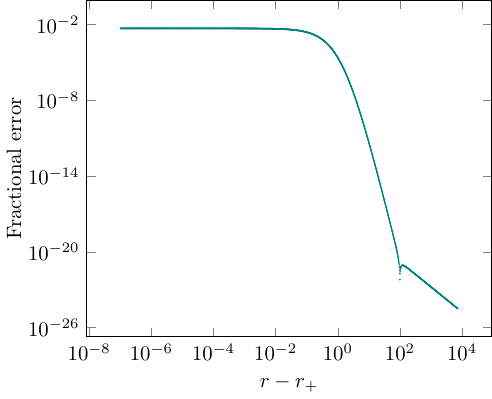}}

\caption{\label{fig:Ideal-gas-radial-error} The fractional error in the source of the Teukolsky equation for a stationary ideal gas cloud when calculated using our decomposition method. For $a=0.9$, $\Lambda=1$, $P=1$, and $\theta=0.01$. The error is less than $0.5\%$ and decreases as one moves away from the horizon as expected. }
\end{figure}

Near the horizon is the most inaccurate zone for our decomposition. This can be seen in Fig.~\ref{fig:Ideal-gas-radial-error}, which plots the radial profile of the error for $a=0.9$ and $\theta=0.01$. Additional precision errors can be encountered near the horizon if insufficient precision is used (less than 32 digits of precision). This is due to the Kinnersley tetrad being singular at the horizon. Hence, when expressing regular quantities using the Kinnersley tetrad near the horizon, cancellations of large numbers are required. This problem could be avoided by using a regular tetrad near the horizon, such as the Hartle--Hawking tetrad~\cite{hawking1972energy, poisson2004absorption}.

\section{Conclusion}\label{sec:Conclusion}

In this paper, we produced a formalism for decomposing the source of the Teukolsky equation into spin-weighted spheroidal harmonics analytically. Our formalism leverages the Held-formalism's~\cite{held1974formalism} ability to extract the $\rho$ and $\rhob$ dependency in background Kerr quantities. The remaining Kerr quantities exhibit a straightforward separation of variables in the Kinnersley tetrad~\cite{kinnersley1969type}. We then used a truncated Fourier series to separate the angular and radial dependency in $\rho$ and $\rhob$. An order $k=8$ Fourier series has a less than $0.5\%$ error for approximating $\Sigma^{-6}$. This error should be sufficiently small for second-order self-force calculations, as the error rapidly decreases as one moves away from the horizon, and second-order effects are small compared to first-order effects. Additionally, the error can be decreased by increasing the order of the expansion.

In section~\ref{sec:IdealGasExample}, we demonstrated our method with a toy example: separating the source of the first-order Teukolsky equation for a stress-energy tensor of a pressure-less ideal gas. The calculation is publicly available in a \textit{Mathematica} notebook~\cite{Teukolsky-source-decomposer-notebook}, which explicitly goes through each step of applying our formalism, as summarised in \cref{fig:Flow-chart}. The calculations confirm the accuracy of our formalism, as we show the error in our source is the same as the error in our expansion for $\rho$ and $\rhob$, as expected.

There is an alternative method for separating the source of the second-order Teukolsky equation. Using the orthogonality of the spin-weighted spheroidal harmonics~\cite{breuer1977some}, as used in Ref.~\cite{Ma:2024qqnm} for a quadratic quasi-normal mode calculation. However, our method provides the efficiency advantage of being analytic, avoiding the need to integrate at each radial point on a grid. This efficiency saving may be crucial for the more involved second-order self-force calculations in Kerr. 

Implementing our method for the first second-order self-force calculations in Kerr is currently in progress. Our formalism will also be used to help calculate second-order (quadratic) quasi-normal mode calculations in Kerr~\cite{loutrel2021second,ripley2021numerical,ioka2007second,nakano2007second, cheung2023nonlinear,mitman2023nonlinearities, Lagos:2022otp,Redondo-Yuste:2023seq,Ma:2024qqnm}.  Additionally, the formalism could be applied to solve Teukolsky equations in theories of gravity beyond general relativity, which generally have extended sources~\cite{hussain2022approach, li2023perturbations,Spiers:2023cva,wagle2023perturbations,d2023ringdowns}.

This paper focuses on applying our method in the frequency domain, but the formalism is also applicable in the time domain. In the time domain, decomposing into spin-weighted spherical harmonics reduces the 2+1 PDE Teukolsky equation into a coupled set of 1+1 PDEs with only nearest and next-to-nearest $\l$-mode coupling~\cite{barack1999late, barack2017time, dolan2013superradiant, long2021time, markakis2023symmetric}. 
Hence, our formalism could help produce the first second-order self-force calculations in the time domain.

\begin{acknowledgments}
AS acknowledges the partial support of a Royal Society University Research Fellowship Enhancement Award and partial support from the STFC Consolidated Grant no. ST/V005596/1. AS would like to thank Adam Pound, Zach Nasipak, and Sam Dolan for helpful discussions. AS would like to thank Laura Sberna, Alexander Grant, and their PRD reviewer for their comments on earlier manuscripts of this work. ChatGPT 3.5 was used to help format equations and figures in this paper. This work makes use of the Black Hole Perturbation Toolkit.

\end{acknowledgments}

\bibliography{bib.bib}

\end{document}